\documentclass[12pt,a4paper]{article}
\usepackage{lmodern}
\usepackage{here}
\usepackage{eurosym}
\usepackage{graphicx}
\usepackage{amsmath}
\usepackage{mathtools}
\usepackage{mathrsfs}
\usepackage{amssymb}
\usepackage{latexsym}
\usepackage{color}
\usepackage{epstopdf}
\usepackage{subcaption}
\usepackage[utf8]{inputenc} 
\usepackage{graphicx}
\usepackage{caption}
\usepackage{braket}
\usepackage[english]{babel}
\usepackage[T1]{fontenc}
\usepackage{float}
\usepackage{cite}
\usepackage[labelsep=space]{caption}

\usepackage{xcolor}
\usepackage{color}
\usepackage{hyperref}
\usepackage{cleveref}
\usepackage[switch]{lineno}
\usepackage[font=footnotesize,labelfont=bf]{caption}
\usepackage[hmargin=2cm,vmargin=2cm,rmargin=2cm, lmargin=2cm]{geometry}
\hypersetup{colorlinks=true,citecolor=red,filecolor=blue,linkcolor=blue,urlcolor=blue}
\DeclareUnicodeCharacter{FB01}{fi}
\DeclareUnicodeCharacter{200B}{}

\def\keywordname{{\bfseries \emph{Keywords}}}%
\def\keywords#1{\par\addvspace\medskipamount{\rightskip=0pt plus1cm
		\def\and{\ifhmode\unskip\nobreak\fi\ $\cdot$}\noindent\keywordname\enspace\ignorespaces#1\par}}

\begin{document}
	\renewcommand{\figurename}{\textbf{Fig.}}
	\renewcommand{\tablename}{\textbf{Table}}
	\newcommand{\un}{\mathds{1}}
	\newcommand{\1}{\mathbbm{1}}
	\newcommand{\vect}[1]{\boldsymbol{#1}}
	\newcommand{\era}{\end{array}}
\newcommand{\beq}{\begin{equation}}
	\newcommand{\eeq}{\end{equation}}
\newcommand{\beqar}{\begin{eqnarray}}
	\newcommand{\eeqar}{\end{eqnarray}}
\newcommand{\lb}{\label}
\thispagestyle{empty}
\baselineskip=18pt
\medskip
\begin{center}
	~~~~~~~~~~~~~~~
	\\
	\vspace{2cm}
	\noindent{{\textbf{Basis-independent coherence and quantum correlations in two dipole-dipole-coupled electrons in double quantum-dot molecules}}}\\
	  
	\vspace{0.3cm}
	
	{\small Zakaria Bouafia}$^{a, }${\footnote{E-mail: \textsf{\href{mailto:zakaria.bouafia1@gmail.com}{zakaria.bouafia1@gmail.com}}}}, {\small Neha Pathania}$^{b, }${\footnote{E-mail: \textsf{\href{mailto:1995npathania@gmail.com}{1995npathania@gmail.com}}}}, and {\small Mostafa Mansour}$^{a, }${\footnote{E-mail: \textsf{\href{mailto:mostafa.mansour.fsac@gmail.com}{mostafa.mansour.fsac@gmail.com}}}} \\
	
	\vspace{0.3cm}
	
	\noindent $^{a}${{\footnotesize LMPHE, Department of Physics, Faculty of Sciences A\"{\i}n Chock,\\ Hassan II University, Casablanca, Morocco}}\\
	
	\noindent $^{b}${{\footnotesize Department of Electronics and Communication, Indian Institute of Technology, Roorkee, Uttarakhand-247667, India}}
\end{center}

\vspace{1cm}

\begin{abstract}
	This work examines the thermal dynamics of basis-independent quantum coherence and correlation-based quantum resources for two dipole-dipole-coupled electrons confined in spatially separated quantum-dot (QD) molecules. Single-dot quantum superpositions and inter-subsystem coherence are characterized by using localized and collective coherence. Quantum correlations between the two double quantum dots are quantified by employing Bures distance entanglement, Local quantum uncertainty (LQU), and local quantum Fisher Information (LQFI). The findings show that dipole-dipole coupling $K$ is the most effective protective parameter, extending the entanglement sudden death temperature,  diminishing the local quantum superpositions and enhancing the collective coherence. The dipole-dipole interaction has also a crucial impact on protecting LQU and LQFI beyond the entanglement sudden death temperature. Coulomb repulsion $J$ reinforces this protection through an independent channel, projecting the thermal state onto the entangled $\{|0_A 1_B\rangle,|1_A 0_B\rangle\}$ subspace; their combined action is required to approach the entanglement maximum, and it enhances collective coherence and extends the temperature range over which LQU and LQFI remain appreciable. Energy detuning $\varepsilon$ can enhance localized coherence but paradoxically accelerates the entanglement sudden death and quenches LQU and LQFI by weakening two-body correlations. Inter-dot tunneling $\Gamma$  enhances local superpositions at low temperature, but it reduces collective coherence, lowers the entanglement sudden death temperature, and disrupts other nonclassical correlations.
\end{abstract}

\vspace{0.3cm}


\vspace{1cm}

\section{Introduction}\label{sec1}

Quantum coherence is a fundamental concept in quantum physics, meaningful even at the single-qubit level, and significantly influences quantum processing tasks. Quantum coherence is determined relative to a selected reference basis. Various methodologies have been developed to assess quantum coherence. These include the skew-information-based measure of coherence~\cite{hu2018quantum}, relative entropy of coherence and $l_1$-norm of coherence~\cite{baumgratz2014quantifying}, correlated coherence~\cite{tan2016unified}, geometric correlated partial coherence~\cite{xiong2019characterizing}, and Hellinger coherence~\cite{hu2016extracting,streltsov2017colloquium,jin2018quantifying}. In quantum thermodynamics, it has been demonstrated that quantum coherence plays a practical role~\cite{rahav2012heat}. Coherence has also been investigated for a fixed basis~\cite{levi2014quantitative,baumgratz2014quantifying,winter2016operational}, for subspaces~\cite{aberg2006quantifying}, for a set of linearly independent states~\cite{killoran2016converting,theurer2017resource}, and as coherence of an entangled state in the Schmidt bases~\cite{pathania2022quantifying}. Coherence and quantum resources are also explored across diverse quantum settings~\cite{benzahra2024dynamics,banouni2024non,bouafia2024decoherence,chouiba2025unveiling,pathania2026dynamics}.

In multipartite systems, entanglement provides the operational advantage that distinguishes quantum from classical charge-qubit architectures~\cite{nielsen2010quantum,walter2016multipartite,beckey2021computable}. Realistic devices operate at finite temperatures, where the Gibbs state progressively mixes the ground-state correlations and degrades entanglement; quantifying this degradation as a function of the Hamiltonian parameters is therefore a prerequisite for charge-qubit design. In general, measuring entanglement in multipartite systems is a challenge. For bipartite systems, however, many measures exist, such as bipartite concurrence~\cite{wootters1998entanglement,wootters2001entanglement}, the Bures distance entanglement~\cite{roga2016,elghaayda2011quantum,bouafia2025comparative}, entanglement negativity~\cite{peres1996separability,horodecki1996separability,eisert1999comparison,vidal2002computable}, and logarithmic negativity~\cite{plenio2005logarithmic,zyczkowski1998volume}. Beyond entanglement, more general measures of nonclassical correlations, such as  LQFI~\cite{kim2018characterizing} and LQU~\cite{girolami2013characterizing}, have attracted considerable attention owing to their close connections with quantum metrology, quantum parameter estimation, and quantum information processing. In semiconductor quantum-dot systems, these quantities have been employed to characterize thermal quantum correlations and to investigate the effects of Coulomb interactions, tunneling processes, and energy detuning on the quantum properties of coupled charge qubits~\cite{aljuaydi2025efficiency,abouie2024entanglement}. Unlike entanglement measures, LQU and LQFI can reveal nonclassical correlations that persist even in regimes where entanglement is strongly suppressed by thermal fluctuations, thereby providing complementary information about the robustness of quantum resources in realistic solid-state architectures.

Semiconductor quantum-dot (QD) molecules are the primary experimental platform for charge-qubit implementations~\cite{filgueiras2020thermal,ait2024nonlocal,dahbi2023effect,ferreira2023thermal,yu2016tunable}: their Hamiltonian parameters are tunable electrostatically, and their fabrication is compatible with standard Si/SiGe processes~\cite{ferreira2023thermal,ait2024nonlocal}. Thermal entanglement and quantum coherence in such systems are sensitive to inter-dot tunneling ($\Gamma$), energy detuning ($\varepsilon$), and Coulomb repulsion ($J$), while dipole--dipole coupling ($K$) between spatially separated molecules introduces an additional inter-qubit channel that modifies the structure of the correlated eigenstates~\cite{ryom2023entanglement,unold2005optical}. Recent studies have addressed thermodynamic cycles~\cite{de2021two,josefsson2020double} in this architecture, but a basis-independent characterization of the full coherence under the simultaneous action of all four Hamiltonian parameters $(\Gamma, \varepsilon, K, J)$ has not been reported. The present work examines the joint effect of the coupling parameter $K$ alongside $\Gamma$, $\varepsilon$, and $J$ in governing the thermal stability of quantum resources in DQD molecules. More precisely, we investigate the impact of the system parameters $(\Gamma, \varepsilon, K, J)$ on the thermal variations of basis-independent quantum coherence, characterized by collective and localized coherence and non-classical correlations, quantified by Bures distance entanglement, LQU, and LQFI between two electrons interacting via dipole-dipole coupling while confined in physically separated quantum-dot molecules.

The paper is organized as follows. Sec.~\ref{sec2} defines the basis-independent coherence measures, Bures distance entanglement, and other nonclassical correlations, LQU and LQFI. The model is presented in Sec.~\ref{sec3}, results are discussed in Sec.~\ref{sec4}, and conclusions follow in Sec.~\ref{sec5}.

\section{Measures of basis-independent quantum coherence and quantum correlations}\label{sec2}

Here, we review the Bures distance entanglement, LQFI, local quantum uncertainty, and basis-independent metrics of coherence.

\subsection{Bures distance entanglement}

For the two quantum density matrices $\rho_1$ and $\rho_2$, the Bures distance~\cite{bures1969extension,uhlmann1976transition,streltsov2010linking} is given by:
\begin{equation}
	\mathcal{B}_d(\rho_1,\rho_2)=\sqrt{2-2\sqrt{F(\rho_1,\rho_2)}},
\end{equation}
where $F(\rho_1,\rho_2)$ denotes the quantum fidelity, defined as:
\begin{equation}
	F(\rho_1,\rho_2)= \mathrm{Tr}\!\left[\sqrt{\sqrt{\rho_2}\,\rho_1\,\sqrt{\rho_2}}\right].
	\label{eq:fidelity}
\end{equation}
The Bures distance entanglement $\mathcal{B}_d(\rho)$ is defined as the minimum Bures distance between $\rho$ and the set of separable states. For two-qubit states, Streltsov~\cite{streltsov2010linking} establishes that the optimal separable state $\sigma^*_{\rm sep}$ satisfies:
\begin{equation}
	F\!\left(\rho,\,\sigma^*_{\rm sep}\right)= \frac{1}{2}+\frac{1}{2}\sqrt{1-\mathcal{C}(\rho)^2},
	\label{eq:fid_conc}
\end{equation}
where $\mathcal{C}(\rho)$ is the concurrence for the two-qubit density matrix $\rho$. Substituting Eq.~(\ref{eq:fid_conc}) into $\mathcal{B}_d(\rho,\sigma^*_{\rm sep})=\sqrt{2-2\sqrt{F(\rho,\sigma^*_{\rm sep})}}$ yields the closed-form expression of the Bures distance entanglement. In this work, we use the normalized expression of the Bures distance entanglement, given as follows:
\begin{equation}\label{Bures01}
	\mathcal{B}_d(\rho)=\frac{1}{\sqrt{2-\sqrt{2}}}[\sqrt{2-\sqrt{2+ 2\sqrt{1-\mathcal{C}(\rho)^2}}}],
\end{equation}
where $\mathcal{C}(\rho)$ is defined in terms of the eigenvalues ${\lambda _n\left( \rho  \right)}$ of the non-Hermitian matrix $\rho\tilde\rho$ as~\cite{wootters1998quantum}:
\begin{align}
	\mathcal{C}\left( \rho  \right) = \max \left\{ {0,2\sqrt{{\lambda _1}\left( \rho  \right)}  - \sum\limits_{n = 1}^4 {\sqrt {{\lambda_n}\left( \rho  \right)} } } \right\},
\end{align}
where the eigenvalues are in descending order and $\tilde\rho={\rm{(}}{\sigma_y} \otimes {\sigma_y}){\rho ^*}{\rm{(}}{\sigma_y} \otimes {\sigma_y})$. 
\subsection{Local quantum uncertainty}

LQU is used to evaluate the skew information-based nonclassical correlations in bipartite quantum states. LQU serves as a significant measure akin to quantum discord and is rooted in the concept of Wigner-Yanase skew information~\cite{wigner1963information,luo2003wigner}. For a bipartite two-qubit state $\rho$, the LQU is defined as~\cite{girolami2013characterizing}:
\begin{equation} \label{U}
	{\cal U}\left(\rho\right)=1-\chi_{\max}(\Lambda),
\end{equation}
where $\chi_{\max}$ represents the largest eigenvalue of the $(3\times3)$ symmetric matrix $\Lambda$, whose elements are given as follows:
\begin{equation}\label{m-elements}
	\Lambda_{ij}=\operatorname{Tr}\left\{\sqrt{\rho}\left(\sigma^i_{A}\otimes \mathbb{I}_{B}\right)\sqrt{\rho}\left(\sigma^j_{A}\otimes\mathbb{I}_{B}\right)\right\},
\end{equation}
with $i,j\in\{x,y,z\}$, $\sigma^{i(j)}_{A}$ are the Pauli operators acting on subsystem $A$. The LQU satisfies the bounds $0\le{\cal U}\left({{\rho}}\right)\le1$, where ${\cal U}\left({{\rho}}\right)=0$ indicates an uncorrelated system state, $0<{\cal U}\left({{\rho}}\right)<1$ implies the presence of non-classical correlations in the system state, and ${\cal U}\left({{\rho}}\right)=1$ corresponds to a maximally correlated system state.

\subsection{Local quantum Fisher information}

In quantum metrology, quantum Fisher information (QFI) is considered a promising indicator for characterizing the precision of parameter estimation in various metrological scenarios~\cite{liu2020quantum}. In this context, local quantum Fisher information (LQFI) is derived from the concept of the QFI, and constitutes a promising measure of discord-type quantum correlations~\cite{kim2018characterizing}. For a bipartite state $(\rho)$, LQFI is defined as follows:
\begin{align}
	\mathcal{Q}(\rho)=\min_{H^{a}}FQ(\rho, H_A),
\end{align} 
with $H_A=H^a\otimes I$. The QFI is given by $FQ(\rho, H_A)$ and the minimization is performed over all local Hamiltonian ${H_A}$ operating on subsystem $A$, where:
\begin{align}\label{QFI}
	FQ\left( {\rho ,{H_A}} \right) = {\rm Tr}\left( {\rho {H_A}^2} \right) - \sum\limits_{i,j} {\frac{{2{\gamma_i}{\gamma_j}}}{{{\gamma_i} + {\gamma_j}}}{{\left| {\left\langle {{\psi _i}} \right|{H_A}\left| {{\psi _j}} \right\rangle } \right|}^2}},
\end{align}
where $\rho=\sum\limits_{i=1}^{4}{{\gamma_i}\left|{{\psi_i}}\right\rangle\langle{\psi_i}|}$ with ${\gamma_i}$ are their corresponding probabilities and $|\psi_i\rangle$ representing the eigenstates. Choosing the local observable $H^a=\vec{\sigma}.\vec{r}$, it can be seen that ${\rm Tr}\left({\rho{H_A}^2}\right)=1$, where $\vec{\sigma}=(\sigma_x, \sigma_y, \sigma_z)$ and $|\vec{r}|=1 $. Therefore, the last term in Eq.~(\ref{QFI}) is given by:
	\begin{equation}
		\sum\limits_{i,j} {\frac{{2{\gamma_i}{\gamma_j}}}{{{\gamma_i} + {\gamma_j}}}{{\left| {\langle {\psi_i}|{H_A}\left| {{\psi_j}} \right\rangle } \right|}^2}} = \sum\limits_{i,j} {\sum\limits_{m,n=1}^3{\frac{{2{\gamma_i}{\gamma_j}}}{{{\gamma_i} + {\gamma_j}}}}} \langle{\psi_i}|{\sigma_m} \otimes \mathbb{I}_B\left|{{\psi_j}} \right\rangle \langle{\psi_j}|{\sigma_n} \otimes \mathbb{I}_B\left|{{\psi_i}} \right\rangle = {{\vec r}^{\dag}}.{\cal W}.{\vec r},
	\end{equation}
where the entries of the matrix ${\cal W}$ ($3 \times 3$) are given as:
\begin{equation}
	{{\cal W}_{mn}} = \sum\limits_{i,j} {\frac{{2{\gamma_i}{\gamma_j}}}{{{\gamma_i} + {\gamma_j}}}\langle {\psi_i}|{\sigma_m} \otimes \mathbb{I}_B\left| {{\psi_j}} \right\rangle \langle {\psi _j}|{\sigma_n} \otimes \mathbb{I}_B\left|{{\psi_i}} \right\rangle}. \label{Mlk}
\end{equation}
To minimize the quantity $FQ\left( {\rho ,{H_A}} \right)$, it is necessary maximizing the quantity ${{\vec r}^\dag }.{\cal W}.{\vec r}$ over all unit vectors ${\vec r}$. Thus, LQFI is given as a function of the eigenvalues of $\cal W$, $\omega_{i(i=1,2,3)}$, by the following explicit formula:
\begin{align}
	\mathcal{Q}(\rho)=1-\text{max}\{\omega_1,\omega_2,\omega_3 \},
\end{align}

\subsection{Localized and collective coherence}

We use the QJSD-based basis-independent coherence framework of Radhakrishnan et al.~\cite{radhakrishnan2019basis,radhakrishnan2016distribution}. For an $n$-qubit state $\eta$ on a $d$-dimensional Hilbert space, the collective coherence $C_C$ and localized coherence $C_L$ are defined as follows. A coherence measure based on the QJSD, defined for a state $\eta$ as:
\begin{equation}\label{CJSD}
	C_{\mathrm{JSD}}(\eta)=\sqrt{\, S\!\left(\tfrac{\eta+\eta_d}{2}\right)-\tfrac{1}{2}\big[S(\eta)+S(\eta_d)\big]},
\end{equation}
where $\eta_d$ denotes the nearest incoherent state obtained by deleting the off-diagonal terms of $\eta$. Since this quantity is inherently basis-dependent, one can instead employ the maximally mixed state $\eta_I=I/d$ as a basis-independent reference, which yields to a basis-independent measure of the localized coherence. This measure isolates the intrinsic quantumness of individual subsystems and is defined as:
\begin{align}\label{CL}
	C_L(\eta)=\sqrt{S\left(\tfrac{\pi_\eta+\eta_I}{2}\right)-\tfrac{1}{2}\big[S(\pi_\eta)+\log_2d\big]}.
\end{align}

The collective coherence, which captures coherence arising from correlations among subsystems, is given by:
\begin{align}\label{CC}
	C_C(\eta)=\sqrt{S\left(\tfrac{\eta+\pi_\eta}{2}\right)-\tfrac{1}{2}\big[S(\eta)+S(\pi_\eta)\big]},
\end{align}
where $\pi_\eta = \eta_1 \otimes \eta_2 \otimes \cdots \otimes \eta_n$ is the tensor product of reduced states.

\section{Two quantum-dot qubits system}\label{sec3}

\begin{figure}[H]
	\centering
	\includegraphics[width=0.7\textwidth]{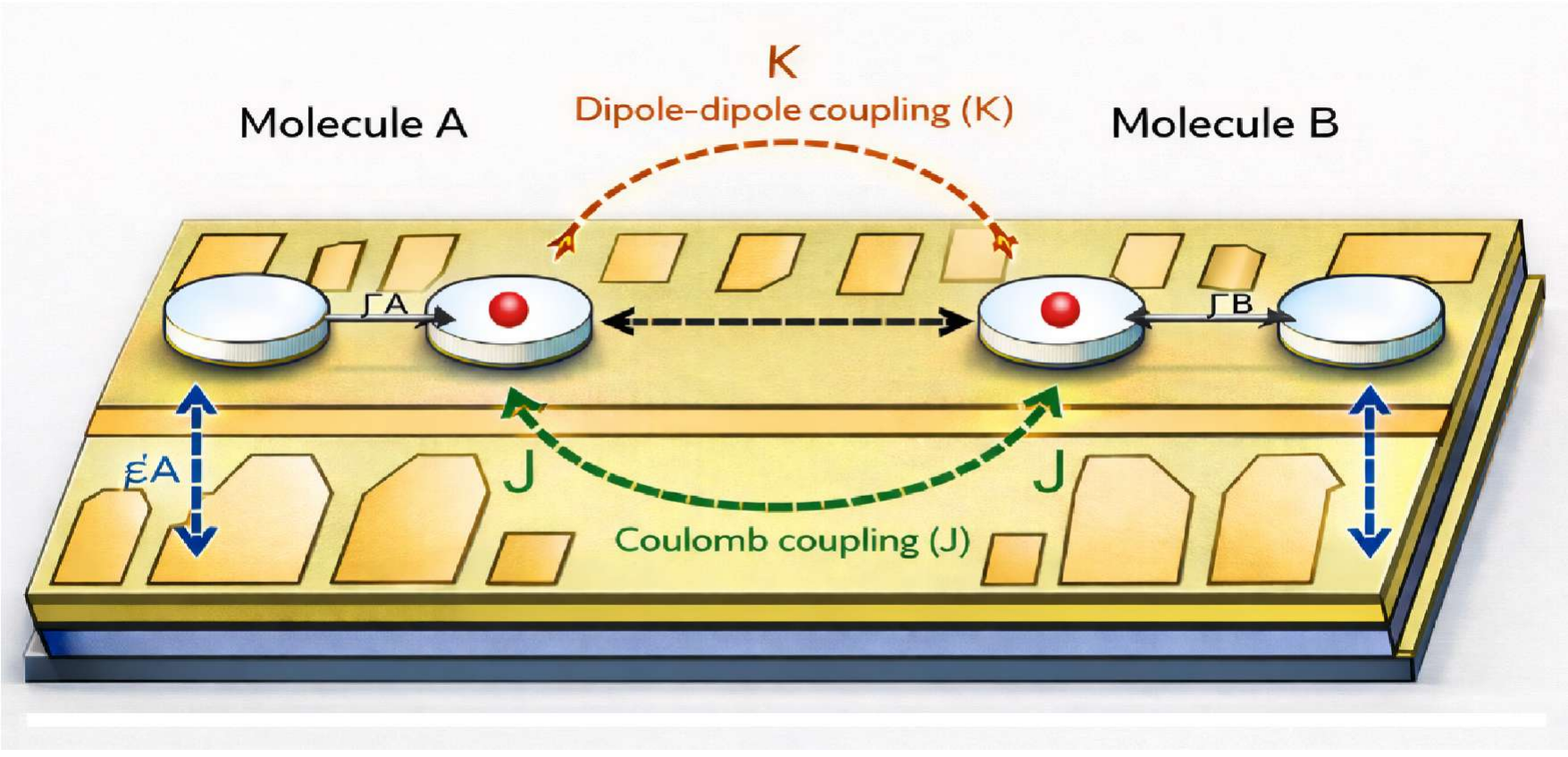}
	\caption{schematic of the two dipole-dipole-coupled DQD molecules with the four coupling parameters $\Gamma_k$, $\varepsilon_k$, $K$, and $J$.}
	\label{mplaf1}
\end{figure}

We consider two spatially separated double-quantum-dot (DQD) molecules, each confining a single electron electrostatically within a tunnel-coupled dot pair~\cite{de2021two,macquarrie2020progress,shinkai2009correlated,barnett2009quantum,wu2000quantum} (Fig.~\ref{mplaf1}). In molecule $k$ ($k=A, B$), the electron position encodes a qubit: $\left|0_k\right\rangle$ and $\left|1_k\right\rangle$ denote localization on the left and right dot, respectively. Intermolecular tunneling between the two molecules is suppressed by design; the qubits are nonetheless coupled through the inter-electronic Coulomb interaction and, additionally, through the dipole--dipole interaction between the two molecules~\cite{ryom2023entanglement,unold2005optical,persaud2020effect,xu2024dipole}.
Within this setup, the Hamiltonian of the system $\hat{H}$ is given by 
\begin{equation}\label{eq:9}
	\hat{H}=\sum_{k=A, B} \frac{\Gamma_k}{2} \hat{\sigma}_k^x+\frac{\varepsilon_k}{2} \hat{\sigma}_k^z+K\left(\hat{\sigma}_A^{+} \hat{\sigma}_B^{-}+\hat{\sigma}_A^{-} \hat{\sigma}_B^{+}\right) \\
	+\lambda_J\left(\hat{I}-\hat{\sigma}_A^z\right) \otimes\left(\hat{I}-\hat{\sigma}_B^z\right).
\end{equation}
The inter-electronic Coulomb repulsion is modeled as $\lambda_J\,(\hat{I}-\hat{\sigma}^z_A)\otimes(\hat{I}-\hat{\sigma}^z_B)$ with $\lambda_J=J/4$, where $J>0$ is the Coulomb coupling strength between the excess electrons. In the basis $\mathcal{B}=\{|0_A0_B\rangle,|0_A1_B\rangle,|1_A0_B\rangle,|1_A1_B\rangle\}$, the matrix form of $\hat{H}$ can be written as
\begin{equation}
	\hat{H}=\left(\begin{array}{cccc}
		\frac{\varepsilon_A+\varepsilon_B}{2} & \frac{1}{2} \Gamma_B & \frac{1}{2} \Gamma_A & 0 \\
		\frac{1}{2} \Gamma_B & \frac{\varepsilon_A-\varepsilon_B}{2} & K & \frac{1}{2} \Gamma_A \\
		\frac{1}{2} \Gamma_A & K & \frac{\varepsilon_B-\varepsilon_A}{2} & \frac{1}{2} \Gamma_B \\
		0 & \frac{1}{2} \Gamma_A & \frac{1}{2} \Gamma_B & J-\frac{\varepsilon_A+\varepsilon_B}{2}
	\end{array}\right).
\end{equation}
Here, $\Gamma_k$ ($k=A, B$) and $\varepsilon_k$ are, respectively, the tunnel and detuning coupling of dot pair $k$, and $\hat{I}$ is the identity matrix. $\hat{\sigma}_k^{x,z}$ are the $k$-qubit Pauli matrices and $\hat{\sigma}_k^{\pm}=\frac{1}{2}(\hat{\sigma}_x \pm i\hat{\sigma}_y)$ the raising and lowering operators. Throughout the numerical study, we set $\Gamma_A=\Gamma_B\equiv\Gamma$ and $\varepsilon_A=\varepsilon_B\equiv\varepsilon$ to reduce the parameter space while preserving the qualitative physics of the symmetric double-dot configuration. All parameters are in energy units with $\hbar=k_B=1$. Mapping to the $Si/SiGe$ platform of Ref.~\cite{macquarrie2020progress}, where $\Gamma\sim 0.5$--$5\,\mu\mathrm{eV}$ and the dilution-refrigerator temperature satisfies $T_{\rm phys}\lesssim 100\,\mathrm{mK}$ ($\approx 8.6\,\mu\mathrm{eV}$), the dimensionless range $T\in[0,10]$ corresponds to $0$--$100\,\mathrm{mK}$~\cite{macquarrie2020progress,shinkai2009correlated}. The finite-temperature state is the Gibbs state:
\begin{equation}\label{eq:11}
	\rho (T)=\frac{\exp(-\hat{H}/k_BT)}{Z}, \qquad
	Z = \operatorname{Tr}\Big(\exp(-\hat{H}/k_BT)\Big).
\end{equation}
The thermal state $\rho(T)$ is obtained by numerically diagonalizing $\hat{H}$ and evaluating Eq.~(\ref{eq:11}). $k_B$ is set to $1$, where the temperature $T$ is measured in units of energy, and all other parameters are dimensionless ratios related to the energy scale. The reduced states $\eta_{A,B}=\mathrm{Tr}_{B,A}(\rho)$ are obtained by partial trace, and the product state $\pi_\eta=\eta_A\otimes\eta_B$ follows. The von Neumann entropy entering Eqs.~(\ref{CJSD})--(\ref{CC}) is $S(\eta)=-\mathrm{Tr}(\eta\log_2\eta)$, with $d = 4$ for the two-qubit system Hilbert space. All quantities are computed as continuous functions of $T\in[0,10]$ in units where the interaction energies are of order unity.

\section{Main results and discussions}\label{sec4}

\begin{figure}[H]
	\centering
	\subfloat[]{%
		\includegraphics[width=.31\textwidth]{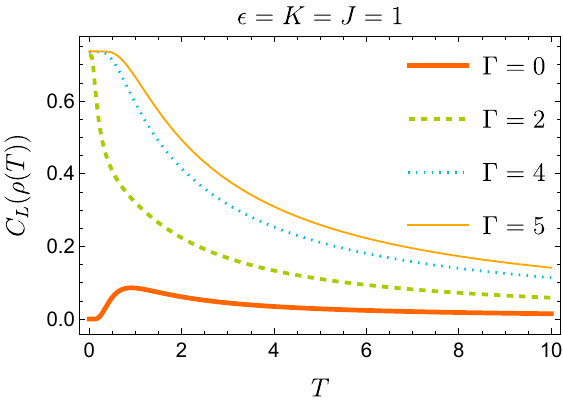}
		\label{2a}}
	\hfill
	\subfloat[]{%
		\includegraphics[width=.31\textwidth]{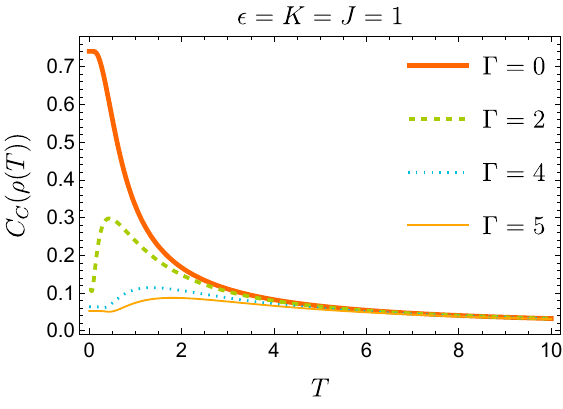}
		\label{2b}}
	\hfill
	\subfloat[]{%
		\includegraphics[width=.31\textwidth]{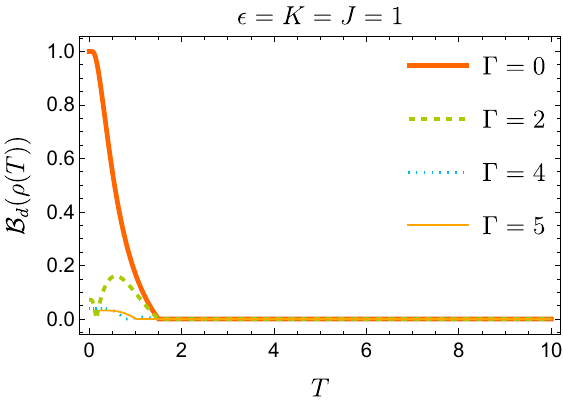}
		\label{2c}}\\
	\subfloat[]{%
		\includegraphics[width=.31\textwidth]{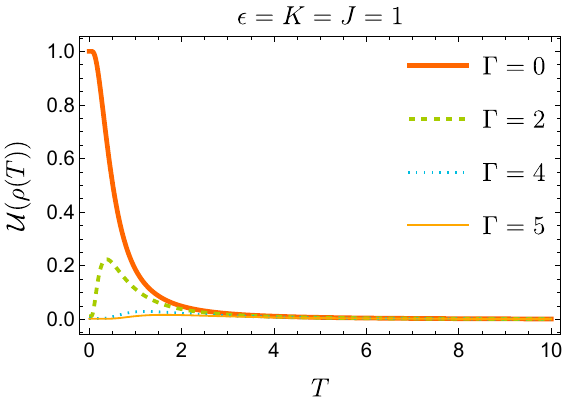}
		\label{2d}}
	\hfill
	\subfloat[]{%
		\includegraphics[width=.31\textwidth]{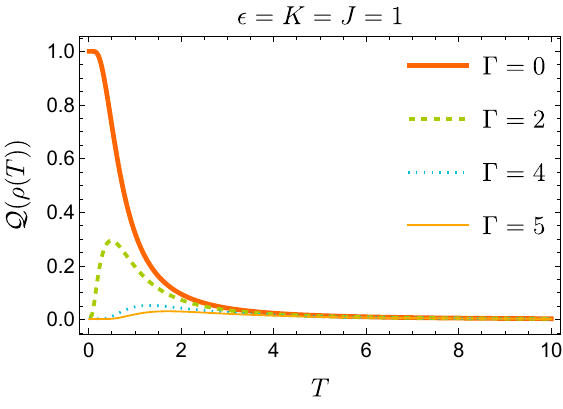}
		\label{2e}}
	\hfill
	\subfloat[]{%
		\includegraphics[width=.31\textwidth]{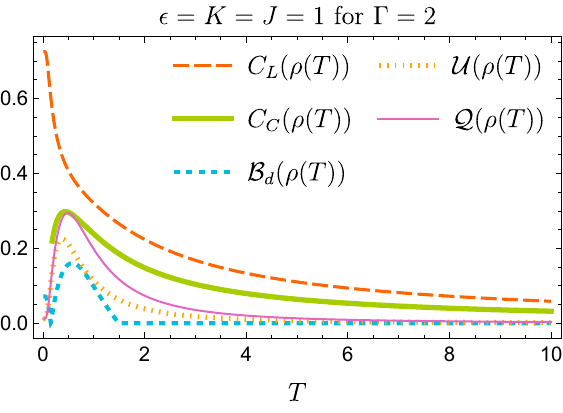}
		\label{2f}}
	\caption{Finite-temperature variation of localized coherence $C_L(\rho(T))$ (\ref{2a}), collective coherence $C_C(\rho(T))$ (\ref{2b}), Bures distance entanglement $\mathcal{B}_d(\rho(T))$ (\ref{2c}), local quantum uncertainty $\mathcal{U}(\rho(T))$ (\ref{2d}), local quantum Fisher information $\mathcal{Q}(\rho(T))$ (\ref{2e}) for different values of $\Gamma$. The remaining Hamiltonian parameters are fixed at $\varepsilon=J=K=1$. Their comparison is provided in (\ref{2f}) for fixed $\Gamma=2$ and $\varepsilon=J=K=1$.}
	\label{fig2}
\end{figure}

Figure~\ref{fig2} presents the temperature dependence of the quantum resources for different values of the inter-dot tunneling strength $\Gamma$, whereas the energy detuning, Coulomb interaction, and dipole-dipole coupling are fixed at $\varepsilon=J=K=1$.
While increasing $\Gamma$ promotes electron delocalization and consequently enhances localized coherence, it simultaneously suppresses collective coherence, entanglement, and information-theoretic quantum correlations. Figure~\ref{2a} shows the temperature dependence of the localized coherence $C_L(\rho(T))$ for different values of the inter-dot tunneling strength $\Gamma$. For $\Gamma=0$, the electrons remain spatially localized and $C_L(\rho(T))$ assumes relatively small values throughout the temperature range $T\in[0,10]$. As $\Gamma$ increases, the electronic states become progressively delocalized across the two quantum dots, leading to a significant enhancement of the localized coherence. Consequently, the low-temperature value of $C_L(\rho(T))$ increases markedly with $\Gamma$, reaching a maximum value of approximately $0.74$ for $\Gamma=5$. With increasing temperature, $C_L(\rho(T))$ gradually decreases due to thermal fluctuations. Nevertheless, stronger tunneling improves the thermal robustness of the localized coherence, allowing it to remain finite over a wider temperature range. These results indicate that inter-dot tunneling promotes local quantum superpositions and enhances their persistence against thermal decoherence. Unlike the localized coherence, the collective coherence $C_C(\rho(T))$ exhibits a markedly different dependence on the inter-dot tunneling strength $\Gamma$. Figure~\ref{2b} illustrates the thermal evolution of the collective coherence $C_C(\rho(T))$ for different values of the inter-dot tunneling strength $\Gamma$.  $C_C(\rho(T))$ decreases significantly as the tunneling strength increases. At $T=0$, the largest value of $C_C(\rho(T))$ is obtained for $\Gamma=0$, where the electrons remain localized, and the Coulomb and dipole-dipole interactions generate strong inter-dot quantum correlations. As the temperature increases, $C_C(\rho(T))$ decreases monotonically for all values of $\Gamma$ due to thermal mixing of the energy eigenstates. Furthermore, increasing $\Gamma$ suppresses the overall magnitude of $C_C(\rho(T))$ throughout the temperature range considered. This behavior can be attributed to the progressive delocalization of the electronic states across the two quantum dots, which weakens the inter-subsystem correlations responsible for collective coherence. Consequently, strong tunneling is detrimental to the preservation of collective quantum coherence. The Bures distance entanglement ($\mathcal{B}_d(\rho (T))$) displayed in Fig.~\ref{2c} is the most sensitive quantity to inter-dot tunneling. In the absence of tunneling ($\Gamma=0$), the system exhibits nearly maximal entanglement at low temperature ($\mathcal{B}_d(\rho (T))=1$) and maintains a finite value up to the entanglement sudden-death temperature ($T^* = 1.5$). Increasing $\Gamma$ rapidly suppresses the low-temperature entanglement and shifts the sudden-death temperature towards smaller values. For example, at $\Gamma=5$, we obtain $T^* = 1$. Physically, strong tunneling delocalizes the electrons across both quantum dots, thereby reducing the effective distinguishability of the two subsystems and weakening the bipartite quantum correlations responsible for entanglement. For $\Gamma\gtrsim 2$, only a small residual entanglement survives at low temperature before being completely destroyed by thermal fluctuations. The LQU $\mathcal{U}(\rho(T))$ and LQFI $\mathcal{Q}(\rho(T))$, shown in Figs.~\ref{2d} and \ref{2e}, exhibit a behavior remarkably similar to that of the Bures distance entanglement. Both quantities attain their largest values for $\Gamma=0$ ($\mathcal{U}(\rho(T))=\mathcal{Q}(\rho(T))=1$) and decrease rapidly as the tunneling strength increases. For moderate and strong tunneling, their magnitudes become strongly suppressed across the entire temperature range, indicating that the quantum correlations accessible through local measurements are highly vulnerable to electron delocalization. The close correspondence between $\mathcal{U}(\rho(T))$, $\mathcal{Q}(\rho(T))$, and $\mathcal{B}_d(\rho(T))$ suggests that these quantities are governed by the same underlying nonclassical correlations generated by the Coulomb and dipole-dipole interactions. The comparative behavior of all quantifiers is summarized in Fig.~\ref{2f}. A clear hierarchy of thermal robustness emerges: $C_L(\rho(T))$ is the most resilient quantum resource, followed by $C_C(\rho(T))$, whereas $\mathcal{U}(\rho(T))$, $\mathcal{Q}(\rho(T))$, and $\mathcal{B}_d(\rho(T))$ decay much more rapidly with temperature.
\begin{figure}[H]
	\centering
	\subfloat[]{%
		\includegraphics[width=.31\textwidth]{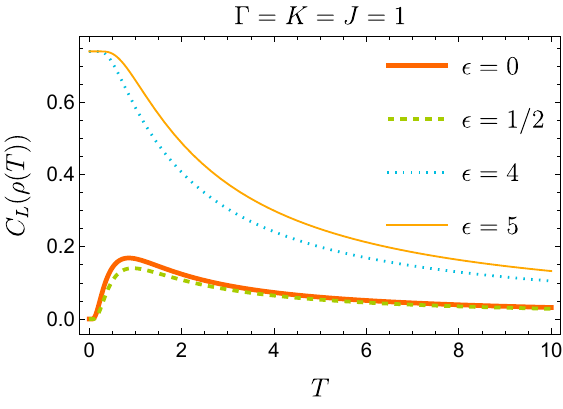}
		\label{3a}}
	\hfill
	\subfloat[]{%
		\includegraphics[width=.31\textwidth]{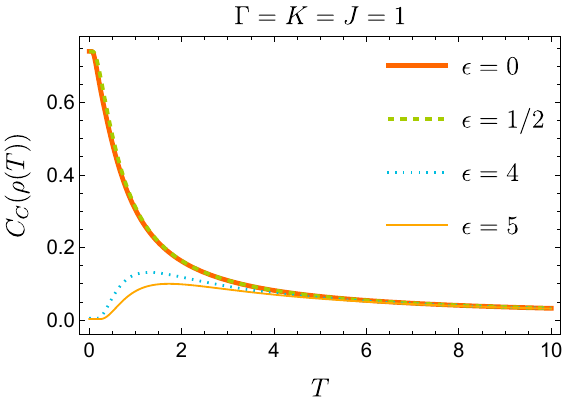}
		\label{3b}}
	\hfill
	\subfloat[]{%
		\includegraphics[width=.31\textwidth]{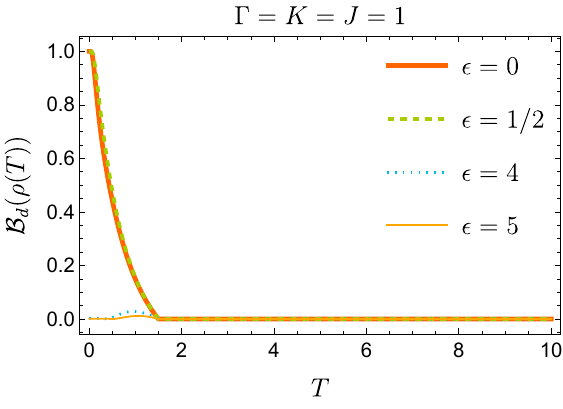}
		\label{3c}}\\
	\subfloat[]{%
		\includegraphics[width=.31\textwidth]{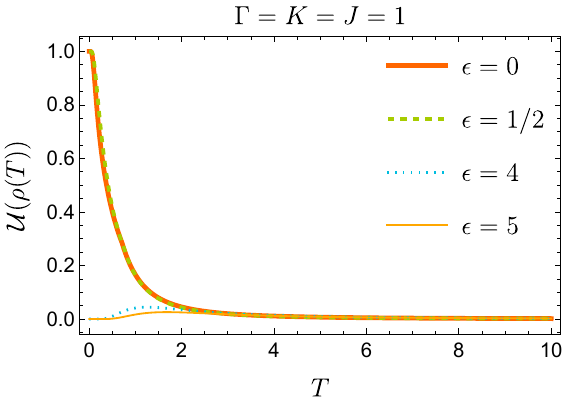}
		\label{3d}}
	\hfill
	\subfloat[]{%
		\includegraphics[width=.31\textwidth]{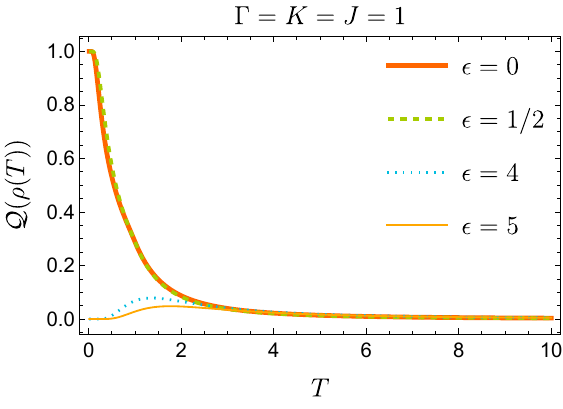}
		\label{3e}}
	\hfill
	\subfloat[]{%
		\includegraphics[width=.31\textwidth]{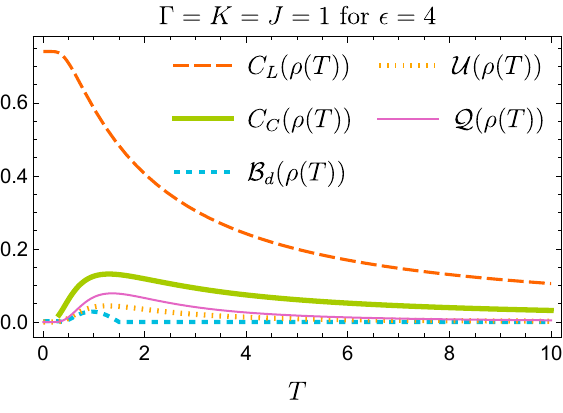}
		\label{3f}}
	\caption{Finite-temperature variation of localized coherence $C_L(\rho(T))$ (\ref{3a}), collective coherence $C_C(\rho(T))$ (\ref{3b}), Bures distance entanglement $\mathcal{B}_d(\rho(T))$ (\ref{3c}), local quantum uncertainty $\mathcal{U}(\rho(T))$ (\ref{3d}), local quantum Fisher information $\mathcal{Q}(\rho(T))$ (\ref{3e}) for some values of the parameter $\varepsilon$. The remaining Hamiltonian parameters are fixed at $\Gamma=J=K=1$. Their comparison is provided in (\ref{3f}) for fixed $\varepsilon=4$ and $\Gamma=J=K=1$.}
	\label{fig3}
\end{figure}
Figure~\ref{fig3} shows the effect of the energy detuning parameter $\varepsilon$ on the thermal behavior of the nonclassical resources for fixed inter-dot tunneling, Coulomb interaction, and dipole-dipole coupling strengths ($\Gamma=J=K=1$). The most pronounced effect of detuning is observed in the localized coherence $C_L(\rho(T))$ shown in Fig.~\ref{3a}. For small detuning ($\varepsilon=0$ and $0.5$), $C_L(\rho(T))$ remains relatively weak, reaching a maximum value of approximately $0.17$ near $T\approx1$ before gradually decreasing with temperature. A qualitatively different behavior emerges for larger detuning values. For $\varepsilon=4$ and $5$, $C_L(\rho(T))$ already assumes a large value of about $0.74$ at low temperatures and remains significantly larger throughout the entire temperature range investigated. Detuning introduces an energy asymmetry between the dots, favoring local superpositions while simultaneously reducing thermal mixing between nearby states. In contrast to the localized coherence, the collective coherence $C_C(\rho(T))$ (Fig.~\ref{3b}) is suppressed by large energy detuning. For $\varepsilon=0$ and $\varepsilon=1/2$, $C_C(\rho(T))$ attains its  highest value of approximately $0.74$ at low temperatures and decays monotonically with increasing temperature. However, for larger detuning strengths ($\varepsilon=4$ and $5$), the low-temperature value of $C_C(\rho(T))$ is strongly reduced and a shallow maximum appears around $T\approx1.5$, after which $C_C(\rho(T))$ gradually decreases. The peak value reaches approximately $0.13$ for $\varepsilon=4$ and $0.10$ for $\varepsilon=5$, which is substantially smaller than that obtained in the weak-detuning regime. While $\varepsilon$ enhances local quantum superpositions by introducing an asymmetry between the quantum dots, it simultaneously weakens the inter-dot coherence shared between the subsystems. As a result, increasing $\varepsilon$ favors localized coherence at the expense of collective coherence. At higher temperatures, thermal fluctuations dominate and drive $C_C(\rho(T))$ towards small values for all detuning strengths. The influence of energy detuning on the Bures distance entanglement $\mathcal{B}_d(\rho(T))$ is shown in Fig.~\ref{3c}. For weak detuning ($\varepsilon=0$ and $1/2$), $\mathcal{B}_d(\rho(T))$ assumes its maximum value at low temperature and decreases rapidly with increasing temperature, vanishing completely at approximately $T^* \approx1.5$. Increasing the detuning strongly suppresses the entanglement. For $\varepsilon=4$ and $5$, at low temperature, entanglement is nearly absent, with only a small thermally induced peak appearing around $T\approx1$. The maximum value of this peak remains below $0.03$, after which $\mathcal{B}_d(\rho(T))$ rapidly vanishes. This behavior indicates that large energy detuning weakens the bipartite quantum correlations between the two quantum dots. A similar trend is observed for the LQU [$\mathcal{U}(\rho(T))$] shown in Fig.~\ref{3d}. For small detuning, $\mathcal{U}(\rho(T))$ starts from its maximum value $\mathcal{U}(\rho(T))=1$ at low temperature and decreases monotonically with increasing temperature. As $\varepsilon$ increases, the magnitude of $\mathcal{U}(\rho(T))$ is strongly reduced and a weak non-monotonic behavior emerges, characterized by a shallow maximum around $T\approx1$. LQFI [$\mathcal{Q}(\rho(T))$], presented in Fig.~\ref{3e}, follows essentially the same qualitative behavior as $\mathcal{U}(\rho(T))$ and $\mathcal{B}_d(\rho(T))$. For $\varepsilon=0$ and $1/2$, $\mathcal{Q}(\rho(T))$ decreases monotonically from its maximum low-temperature value, whereas for large detuning, only a small finite-temperature peak survives before thermal fluctuations suppress the quantum Fisher information completely. Since $\mathcal{Q}(\rho(T))$ quantifies the sensitivity of the state to local parameter estimation, these results indicate that strong detuning significantly reduces the metrological utility of the quantum state by diminishing the underlying quantum correlations. The comparative behavior of all quantum resources is summarized in Fig.~\ref{3f} for $\varepsilon=4$ and $\Gamma= K=J=1$. $C_L(\rho(T))$ is the most resilient quantity, retaining a substantial value throughout the entire temperature range considered. $C_C(\rho(T))$ exhibits intermediate robustness, remaining finite even after the correlation-based measures have become strongly suppressed. In contrast, $\mathcal{B}_d(\rho(T))$, $\mathcal{Q}(\rho(T))$, and $\mathcal{U}(\rho(T))$, rapidly decay and remain confined to the low-temperature regime. Notably, $\mathcal{Q}(\rho(T))$ and $\mathcal{U}(\rho(T))$ survive slightly longer than the entanglement, indicating that non-classical correlations persist even after the complete disappearance of bipartite entanglement. Overall, the results demonstrate that energy detuning preferentially preserves local coherence while strongly suppressing the nonlocal quantum correlations responsible for entanglement, local uncertainty, and quantum-enhanced metrological sensitivity.
\begin{figure}[H]
	\centering
	\subfloat[]{%
		\includegraphics[width=.31\textwidth]{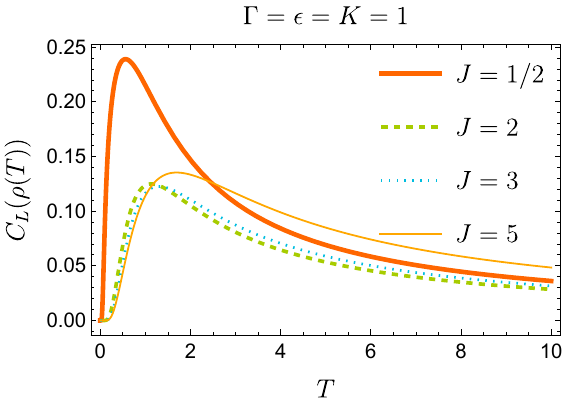}
		\label{4a}}
	\hfill
	\subfloat[]{%
		\includegraphics[width=.31\textwidth]{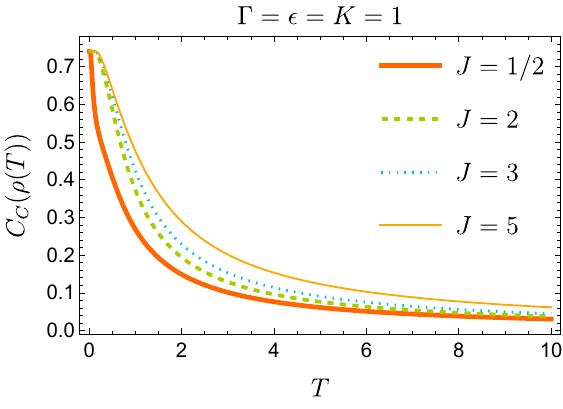}
		\label{4b}}
	\hfill
	\subfloat[]{%
		\includegraphics[width=.31\textwidth]{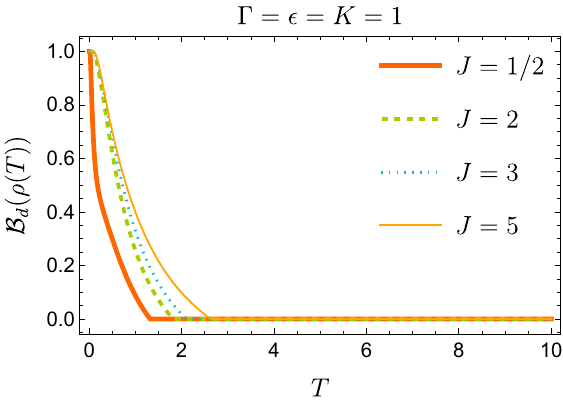}
		\label{4c}}\\
	\subfloat[]{%
		\includegraphics[width=.31\textwidth]{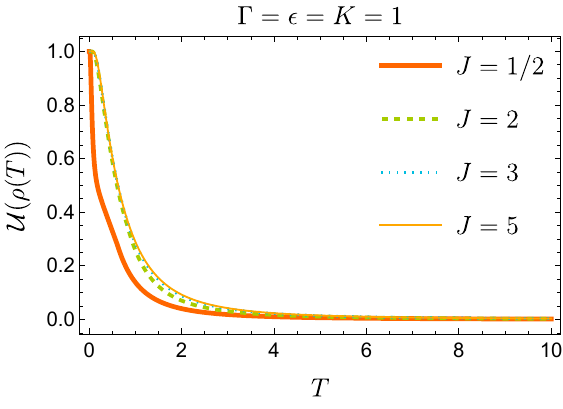}
		\label{4d}}
	\hfill
	\subfloat[]{%
		\includegraphics[width=.31\textwidth]{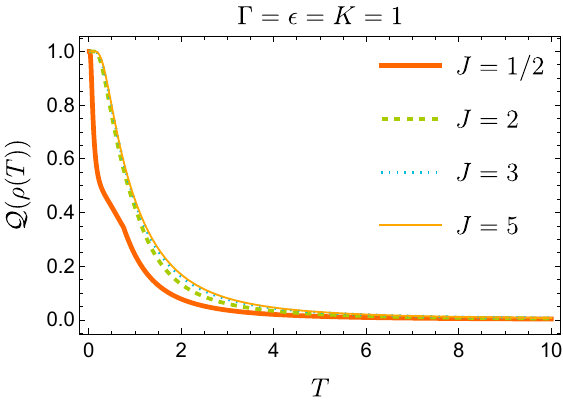}
		\label{4e}}
	\hfill
	\subfloat[]{%
		\includegraphics[width=.31\textwidth]{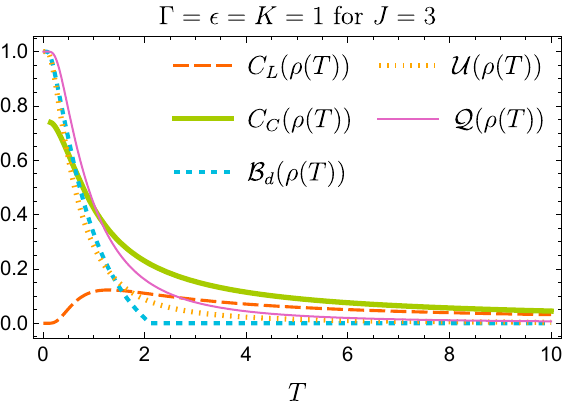}
		\label{4f}}
	\caption{Finite-temperature variation of localized coherence $C_L(\rho(T))$ (\ref{4a}), collective coherence $C_C(\rho(T))$ (\ref{4b}), Bures distance entanglement $\mathcal{B}_d(\rho(T))$ (\ref{4c}), local quantum uncertainty $\mathcal{U}(\rho(T))$ (\ref{4d}), local quantum Fisher information $\mathcal{Q}(\rho(T))$ (\ref{4e}) for some fixed values of the coulomb-coupling strength $J$. The other parameters are fixed at $\Gamma=\varepsilon=K=1$. Their comparison is provided in (\ref{4f}) for fixed $J=3$ and $\Gamma=\varepsilon=K=1$.}
	\label{fig4}
\end{figure}
Figure~\ref{fig4} shows the effect of the Coulomb coupling $J$ between the excess electrons, keeping the other parameters constant at $\Gamma=\varepsilon=K=1$. This interaction raises the energy of the doubly occupied state $|1_A 1_B\rangle$ relative to the other states, penalizing configurations in which both electrons occupy their right-hand dots simultaneously. Increasing $J$ enhances all coherence measures at every temperature; $J$ suppresses $|1_A 1_B\rangle$ energetically, projecting the thermal state toward the $\{|0_A 1_B\rangle, |1_A 0_B\rangle\}$ subspace. $K$, by contrast, directly couples these two states in the Hamiltonian, generating eigenstates that are entangled at the Hamiltonian level. In contrast to the effects of inter-dot tunneling and energy detuning, increasing the Coulomb interaction J generally enhances collective coherence, entanglement, LQU, and LQFI, while simultaneously improving their thermal robustness. These results indicate that the Coulomb interaction plays a positive role in establishing and preserving inter-dot quantum correlations. $C_L(\rho(T))$ (Fig.~\ref{4a}) exhibits a non-monotonic dependence on the Coulomb interaction strength $J$. For all values of $J$, $C_L(\rho(T))$ initially increases with temperature, reaches a maximum, and subsequently decreases due to thermal decoherence. The largest peak value, approximately $0.24$, is obtained for $J=1/2$ near $T \approx 0.5$, whereas larger Coulomb interactions lead to lower peak values. However, increasing $J$ improves the thermal persistence of $C_L(\rho(T))$, as evidenced by the larger values maintained at intermediate and high temperatures. Unlike the localized coherence, the collective coherence $C_C(\rho(T))$ is strongly enhanced by the Coulomb interaction strength $J$. As shown in Fig.~\ref{4b}, all curves start from nearly the same low-temperature value, $C_C(\rho(T)) \approx 0.74$, indicating that the ground-state coherence is largely insensitive to $J$ within the parameter range considered. However, as we increase the temperature, significant differences emerge. Larger values of $J$ lead to a noticeably slower decay of $C_C(\rho(T))$, resulting in substantially higher coherence at intermediate and high temperatures. A similar trend is observed for $\mathcal{B}_d(\rho(T))$ shown in Fig.~\ref{4c}. Although all curves begin with nearly maximal entanglement at low temperature, the thermal robustness of the entangled state increases significantly with $J$. For $J=1/2$, the entanglement vanishes at approximately $T^*\approx1.3$, whereas for $J=5$ it survives up to nearly $T^*\approx2.65$. Thus, increasing the Coulomb interaction extends the entanglement sudden-death temperature by just over a factor of two. Physically, the Coulomb interaction favors the single-excitation subspace ${|0_A1_B\rangle,|1_A0_B\rangle}$, which supports the strongest quantum correlations and therefore stabilizes the entangled state against thermal fluctuations. LQU and LQFI displayed in Fig. \ref{4d} and Fig. \ref{4e}, respectively, closely follow the behavior shown in ~\ref{4c}. In both cases, increasing $J$ slows the thermal decay and shifts the onset of strong suppression towards higher temperatures. The similarity between $\mathcal{U}(\rho(T))$, $\mathcal{Q}(\rho(T))$, and $\mathcal{B}_d(\rho(T))$ indicates that these quantities are governed by the same underlying inter-dot quantum correlations. Furthermore, unlike entanglement, both $\mathcal{U}(\rho(T))$ and $\mathcal{Q}(\rho(T))$ remain finite beyond the entanglement sudden-death temperature, demonstrating that nonclassical correlations and metrologically useful quantum resources persist even after bipartite entanglement has completely vanished. Fig. ~\ref{4f} summarizes the relative thermal robustness of the different quantifiers for $J=3$ and $\Gamma=\varepsilon=K=1$. A clear hierarchy of thermal robustness is observed. While $\mathcal{B}_d(\rho(T))$ exhibits the fastest decay and vanishes completely near $T\approx2.1$, both $\mathcal{U}(\rho(T))$ and $\mathcal{Q}(\rho(T))$ remain finite over a broader temperature range. $C_C(\rho(T))$ is even more resilient, whereas $C_L(\rho(T))$ becomes the dominant resource at elevated temperatures. These results demonstrate that the Coulomb interaction preferentially preserves correlation-based quantum resources and substantially improves their thermal stability, making $J$ an effective, though secondary, protective parameter relative to $K$ (see Fig. \ref{fig5}).
\begin{figure}[H]
	\centering
	\subfloat[]{%
		\includegraphics[width=.31\textwidth]{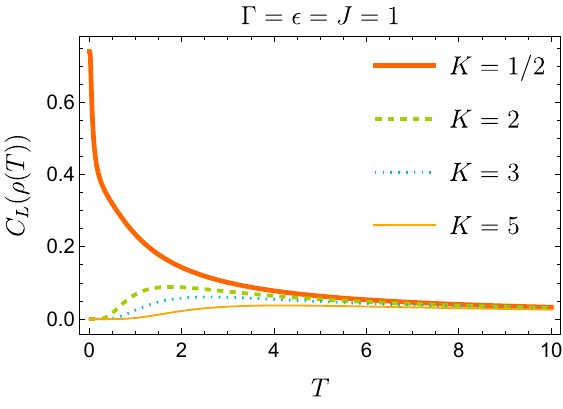}
		\label{5a}}
	\hfill
	\subfloat[]{%
		\includegraphics[width=.31\textwidth]{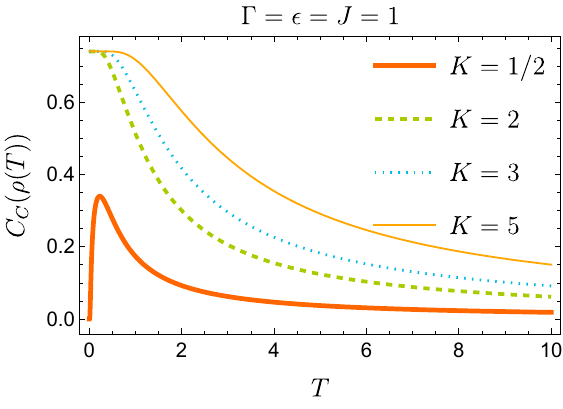}
		\label{5b}}
	\hfill
	\subfloat[]{%
		\includegraphics[width=.31\textwidth]{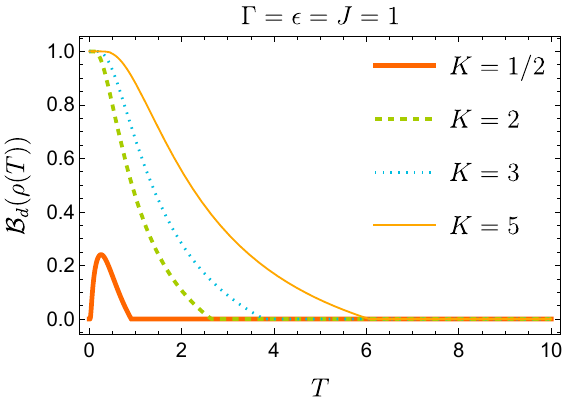}
		\label{5c}}\\
	\subfloat[]{%
		\includegraphics[width=.31\textwidth]{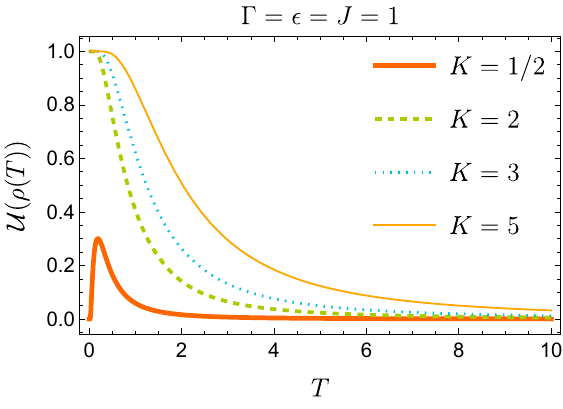}
		\label{5d}}
	\hfill
	\subfloat[]{%
		\includegraphics[width=.31\textwidth]{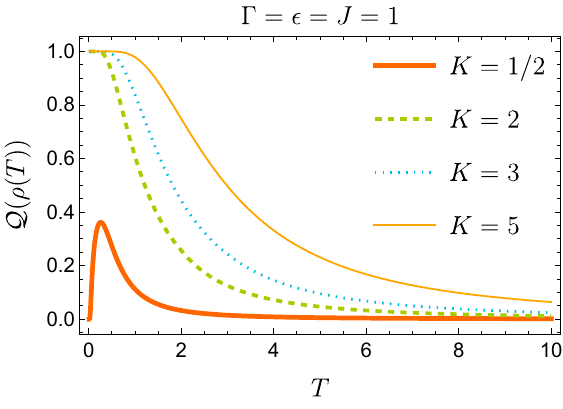}
		\label{5e}}
	\hfill
	\subfloat[]{%
		\includegraphics[width=.31\textwidth]{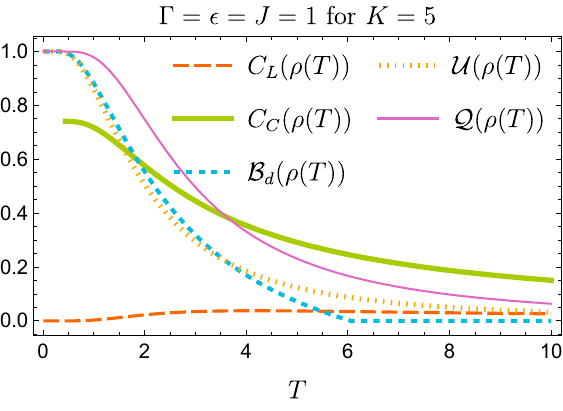}
		\label{5f}}
	\caption{Finite-temperature variation of localized coherence $C_L(\rho(T))$ (\ref{5a}), collective coherence $C_C(\rho(T))$ (\ref{5b}), Bures distance entanglement $\mathcal{B}_d(\rho(T))$ (\ref{5c}), local quantum uncertainty $\mathcal{U}(\rho(T))$ (\ref{5d}), local quantum Fisher information $\mathcal{Q}(\rho(T))$ (\ref{5e}) for different values of the dipole-dipole-coupling strength $K$. The remaining Hamiltonian parameters are fixed at $\Gamma=\varepsilon=J=1$. Their comparison is provided in (\ref{5f}) for fixed $K=5$ and $\Gamma=\varepsilon=J=1$.}
	\label{fig5}
\end{figure}
Figure~\ref{fig5} presents the dependence of coherence and quantum correlations on the dipole-dipole interaction strength $K$ between electrons in different quantum dot molecules. In all panels, increasing $K$ from $1/2$ to $5$ enhances coherence at all temperatures. $C_L(\rho(T))$ exhibits a markedly different dependence on the dipole-dipole coupling strength $K$ compared to the other quantum resources. As shown in Fig.~\ref{5a}, $C_L(\rho(T))$ attains its largest value for weak coupling ($K=1/2$), reaching approximately $0.74$ at low temperatures before decreasing monotonically with increasing temperature. As the dipole-dipole interaction strength increases, the magnitude of $C_L(\rho(T))$ is significantly reduced over the entire temperature range considered. For $K=5$, the localized coherence remains below $0.04$ even at its maximum value. This suppression indicates that strong dipole-dipole interactions diminish the local quantum superpositions associated with individual quantum dots. In sharp contrast, the collective coherence $C_C(\rho(T))$ is strongly enhanced by the dipole-dipole interaction. For weak coupling ($K=1/2$), $C_C(\rho(T))$ exhibits a small thermal peak of approximately $0.34$ before rapidly decaying with temperature. Increasing $K$ enhances both the magnitude and thermal robustness of $C_C(\rho(T))$. For $K=5$, the collective coherence reaches a value close to $0.74$ at low temperatures and remains appreciable even at $T=10$. This behavior originates from the fact that the dipole-dipole interaction directly couples the states $|0_A1_B\rangle$ and $|1_A0_B\rangle$ in the Hamiltonian. Consequently, the energy eigenstates become coherent superpositions of the two single-excitation configurations, generating strong inter-dot correlations even before thermal effects are taken into account. As $K$ increases, coherence is progressively transferred from local degrees of freedom into collective quantum correlations, leading to the suppression of $C_L(\rho(T))$ and the simultaneous enhancement of $C_C(\rho(T))$. The effect of the dipole-dipole interaction is more pronounced for $\mathcal{B}_d(\rho(T))$, shown in Fig.~\ref{5c}. For weak coupling ($K=1/2$), the entanglement is strongly suppressed and survives only within a narrow low-temperature region, vanishing completely at approximately $T^*\approx0.9$. As the dipole-dipole interaction strength increases, both the magnitude and thermal robustness of the entanglement increase dramatically. For $K=5$, $\mathcal{B}_d(\rho(T))$ remains close to its maximum value at low temperatures and persists up to approximately $T^*\approx6$, representing nearly an order-of-magnitude enhancement of the entanglement survival temperature. This strong dependence reflects the fact that $K$ directly entangles the two qubits at the eigenstate level: by mixing the $|0_A 1_B\rangle$,~$|1_A 0_B\rangle$ subspace with the rest of the Hilbert space, it generates eigenstates that are already correlated at the level of the bare Hamiltonian, so thermal occupation of these states naturally preserves entanglement up to temperatures far beyond what Coulomb repulsion or detuning alone can sustain. LQU and LQFI, displayed in Figs.~\ref{5d} and \ref{5e}, exhibit a behavior similar to that of the $\mathcal{B}_d(\rho(T))$. Increasing $K$ significantly slows their thermal decay and extends the temperature range over which they remain appreciable. For weak dipole-dipole coupling, both quantities rapidly decrease to negligible values. In contrast, for $K=5$, $\mathcal{U}(\rho(T))$ and $\mathcal{Q}(\rho(T))$ remain finite over a broad temperature interval, demonstrating that the dipole-dipole interaction not only preserves entanglement but also protects other nonclassical correlations. Figure~\ref{5f} summarizes the comparative behavior of the investigated quantum resources for $K=5$. A clear separation between local and correlation-based quantum resources is observed. While $C_L(\rho(T))$ remains comparatively small throughout the temperature range, the collective coherence $C_C(\rho(T))$, $\mathcal{B}_d(\rho(T))$, $\mathcal{U}(\rho(T))$, and $\mathcal{Q}(\rho(T))$ attain substantially larger values and exhibit enhanced thermal robustness. In particular, $C_C(\rho(T))$ emerges as the most persistent resource, remaining finite even at the highest temperatures considered. These results demonstrate that the dipole-dipole interaction effectively constitutes the more efficient mechanism for protecting quantum resources against thermal degradation among all the parameters investigated in this work.
\begin{figure}[H]
	\centering
	\subfloat[]{%
		\includegraphics[width=.31\textwidth]{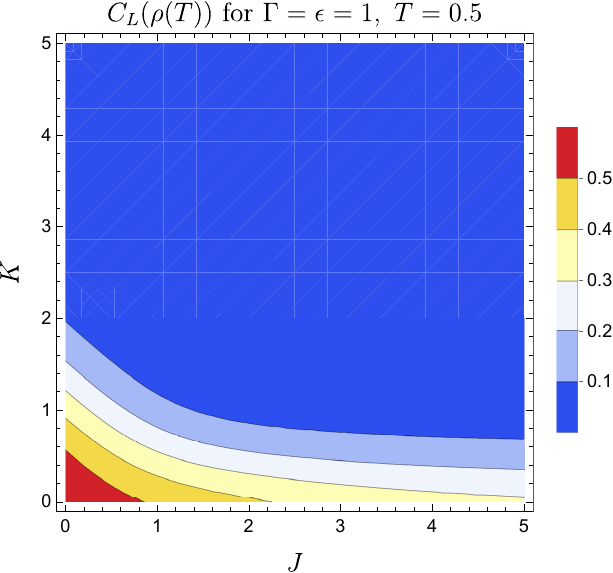}
		\label{6a}}
	\hfill
	\subfloat[]{%
		\includegraphics[width=.31\textwidth]{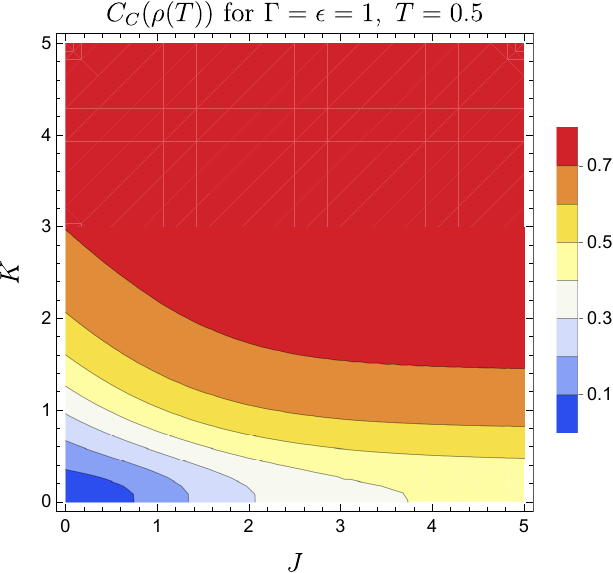}
		\label{6b}}
	\hfill
	\subfloat[]{%
		\includegraphics[width=.31\textwidth]{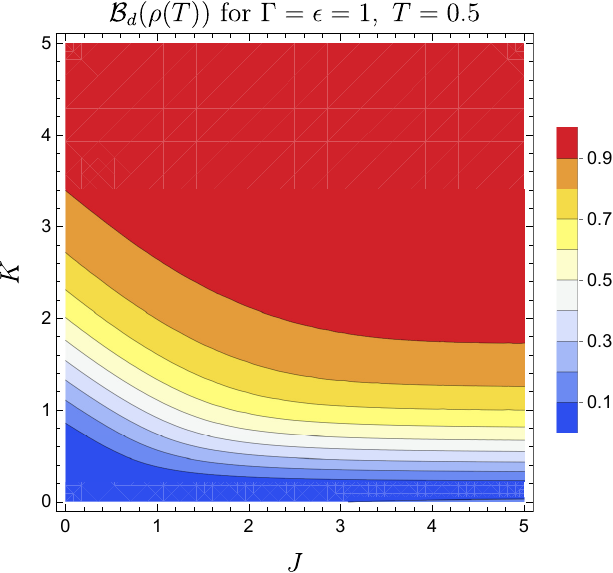}
		\label{6c}}\\
	\subfloat[]{%
		\includegraphics[width=.31\textwidth]{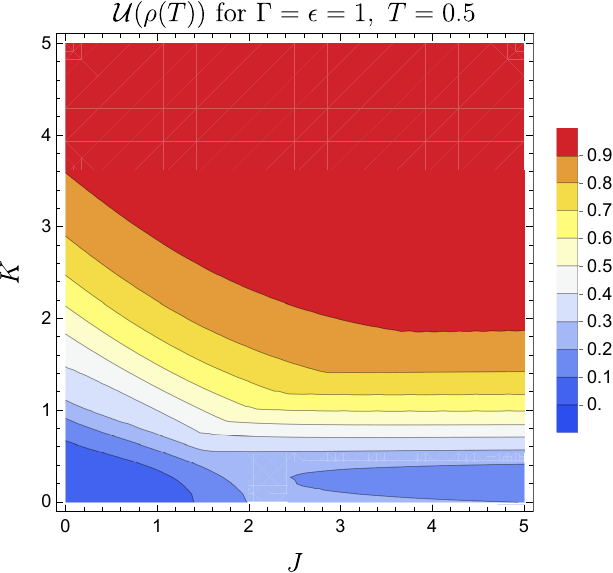}
		\label{6d}}
	\hspace{0.7cm}
	\subfloat[]{%
		\includegraphics[width=.31\textwidth]{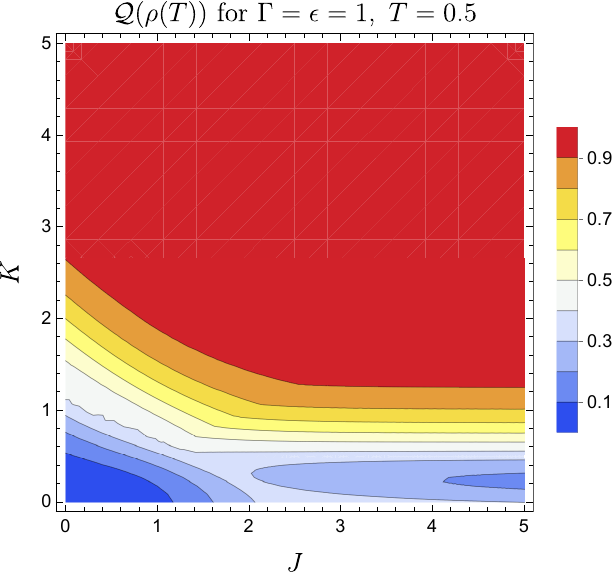}
		\label{6e}}
	\caption{Contour plots of the localized coherence $C_L(\rho(T))$ (\ref{6a}), collective coherence $C_C(\rho(T))$ (\ref{6b}), Bures distance entanglement $\mathcal{B}_d(\rho(T))$ (\ref{6c}), local quantum uncertainty $\mathcal{U}(\rho(T))$ (\ref{6d}), and local quantum Fisher information $\mathcal{Q}(\rho(T))$ (\ref{6e}) in the $(J,K)$ parameter space for fixed $\Gamma=\varepsilon=1$ and $T=0.5$.}
	\label{fig6}
\end{figure}
Figure~\ref{fig6} shows $C_L(\rho(T))$, $C_C(\rho(T))$, $B_d(\rho(T))$, $\mathcal{U}(\rho(T))$, and $\mathcal{Q}(\rho(T))$ as functions of $J$ and $K$ at a specific temperature $T=0.5$ and $\Gamma=\varepsilon=1$. The localized coherence $C_L(\rho(T))$ (Fig.~\ref{6a}) is nearly insensitive to $J$ across the full parameter range, confirming that Coulomb repulsion does not generate single-dot superpositions; its variation with $K$ follows the mechanism discussed in connection with Fig.~\ref{5a}. The collective coherence $C_C(\rho(T))$ (Fig.~\ref{6b}) saturates beyond $J \approx 2$: at $T=0.5$, this value of $J$ is already sufficient to fully suppress the doubly occupied state $|1_A 1_B\rangle$ energetically, so any further increase of $J$ yields no additional inter-subsystem correlation and $C_C(\rho(T))$ is then governed almost entirely by $K$.  The entanglement map (Fig.~\ref{6c}) reveals approximately hyperbolic $B_d(\rho(T))$ contours near the origin, reflecting the fact, that at small coupling strengths, $J$ and $K$ contribute comparably to entanglement protection; as $K$ increases beyond $K \gtrsim 3$, the contours become nearly horizontal, indicating that $B_d(\rho(T))$ is then driven almost exclusively by $K$, with $J$ playing only a secondary stabilizing role. The corner values obtained numerically at $\Gamma=\varepsilon=1$ and $T=0.5$ are: $B_d(\rho(T))\approx0.1$ at $(J=0, K=0)$; $B_d(\rho(T))\approx0.9$ at $(J=0, K=5)$; $B_d(\rho(T))\approx0.1$ at $(J=5, K=0)$; and $B_d(\rho(T))\approx0.9$ at $(J=5, K=5)$. Since $\mathcal{B}_d(\rho(T))\approx0.9$ is already reached at $(J=0,\,K=5)$, the near-maximal entanglement in this corner regime is driven almost entirely by $K$, with $J$ playing only a marginal role, ~consistent with the nearly horizontal contours for $K\gtrsim3$ noted above. The contour maps of the LQU and LQFI, shown in Figs.~\ref{6d} and~\ref{6e}, display behavior closely resembling that of the Bures distance entanglement. In both cases, the contour lines are approximately hyperbolic near the origin, indicating that for weak interactions, the Coulomb and dipole-dipole couplings contribute comparably to the preservation of nonclassical correlations. As the dipole-dipole interaction strength increases, however, the contours progressively flatten and become nearly horizontal for $K\gtrsim3$, revealing that the quantum resources are then governed predominantly by $K$, while the influence of $J$ becomes secondary. This behavior confirms that the dipole-dipole interaction is the principal mechanism responsible for generating and protecting quantum correlations in the system. Unlike the Bures-distance entanglement, neither $\mathcal{U}(\rho(T))$ nor $\mathcal{Q}(\rho(T))$ exhibits a sudden disappearance within the parameter range considered. Instead, both quantities remain finite even in regions where the entanglement is relatively weak, demonstrating that non-classical correlations can survive beyond the regime of strong bipartite entanglement. The highest values of $\mathcal{U}(\rho(T))$ and $\mathcal{Q}(\rho(T))$ are obtained in the upper-right corner of the parameter space, where large Coulomb and dipole-dipole couplings act simultaneously. This confirms that the two interactions protect quantum resources through complementary physical mechanisms: Coulomb repulsion energetically suppresses unfavorable charge configurations, whereas the dipole-dipole interaction generates entangled eigenstates that remain robust against thermal noise.

\section{Conclusion}\label{sec5}

We have investigated the thermal behavior of correlation-based quantum resources and basis-independent quantum coherence in a system of two electrons interacting via dipole-dipole coupling while confined in physically separated quantum-dot molecules. Our results reveal that the four Hamiltonian parameters influence coherence and correlation-based quantum resources in fundamentally different ways. The inter-dot tunneling strength $\Gamma$ predominantly enhances localized coherence by promoting electron delocalization between the quantum dots. However, this delocalization weakens collective coherence, Bures distance entanglement, LQU, and LQFI, reducing the entanglement sudden-death temperature from $T^*\approx1.5$ at $\Gamma=0$ to approximately $T^*\approx1.0$ at $\Gamma=5$. Energy detuning $\varepsilon$ exhibits a similar competition between local and nonlocal quantum resources. Increasing $\varepsilon$ significantly enhances localized coherence while suppressing collective coherence and all correlation-based measures, indicating that the energetic asymmetry favors local quantum superpositions at the expense of inter-dot quantum correlations and, consequently, accelerates the disappearance of entanglement. The Coulomb interaction $J$ plays a constructive role in preserving quantum correlations. By energetically suppressing double occupation and favoring the entangled single-excitation subspace, it enhances collective coherence, entanglement, LQU, and LQFI while extending the entanglement sudden-death temperature from approximately $T^*\approx1.3$ at $J=1/2$ to $T^*\approx2.65$ at $J=5$. Nevertheless, its influence on localized coherence remains comparatively weak. Among all the parameters investigated, the dipole-dipole interaction $K$ emerges as the more effective mechanism for protecting non-classical quantum resources. Increasing $K$ transfers coherence from localized to collective degrees of freedom, leading to a remarkable enhancement of collective coherence, entanglement, LQU, and LQFI. In particular, the entanglement sudden-death temperature increases from approximately $T^*\approx0.9$ at $K=1/2$ to nearly $T^*\approx6$ at $K=5$, demonstrating the exceptional robustness imparted by the generation of intrinsically entangled eigenstates.  For a thermally robust charge-qubit operation, $K$ and $J$ should be maximized jointly, while $\Gamma$ and $\varepsilon$ should be minimized. The complementary roles of the Coulomb and dipole-dipole interactions, together with the competing effects of tunneling and energy detuning, provide valuable physical insight into the interplay between localized coherence, collective coherence, and nonclassical correlations, offering useful guidelines for the design of thermally robust solid-state charge-qubit architectures.

\bibliography{references}
\bibliographystyle{ieeetr}

\end{document}